\documentclass[journal,draftcls,onecolumn,12pt,twoside]{IEEEtranTCOM}
\normalsize

\ifCLASSINFOpdf
\else
\fi

\usepackage{cite}
\usepackage{url}

\usepackage[cmex10]{amsmath}
\usepackage{amsfonts}
\usepackage{algorithmic}
\usepackage{algorithm}
\usepackage{graphicx}
\usepackage{subfigure}
\usepackage{bm}
\usepackage{times,color,amssymb,epsfig,psfrag,stfloats,enumerate}

\begin{document}
%
\title{Base Station Sleeping and Resource \\Allocation in Renewable Energy Powered \\Cellular Networks}


\author{Jie Gong,~\IEEEmembership{Student Member,~IEEE}, John S.~Thompson,~\IEEEmembership{Member,~IEEE}, Sheng Zhou,~\IEEEmembership{Member,~IEEE}, Zhisheng Niu,~\IEEEmembership{Fellow,~IEEE}
\thanks{J.~Gong, S.~Zhou and Z.~Niu are with Tsinghua National Laboratory for Information Science and Technology, Department of Electronic Engineering, Tsinghua University, Beijing 100084, P.~R.~China. Email: gongj08@mails.tsinghua.edu.cn, \{sheng.zhou, niuzhs\}@tsinghua.edu.cn.}
\thanks{J.~S.~Thompson is with the Institute of Digital Communications, School of Engineering, University of Edinburgh, Edinburgh EH9 3JL, UK. Email: John.Thompson@ed.ac.uk.}
\thanks{This work is sponsored in part by the National Science Foundation of China (NSFC) under grant No.~61201191, the National Basic Research Program of China (973 Program: 2012CB316001), the Distinguished Young Scholars of NSFC under grant No.~60925002, the Creative Research Groups of NSFC under grant No.~61021001, and Hitachi R\&D Headquarter.}
\thanks{John Thompson acknowledges part funding of this work by EPSRC Grant EP/J015180/1.}
\thanks{Part of the work has been accepted to appear at IEEE VTC Spring 2013.}

}

\markboth{IEEE Transactions on Wireless Communications}%
{Submitted paper}

\maketitle

\begin{abstract}
We consider energy-efficient wireless resource management in cellular networks where BSs are equipped with energy harvesting devices, using statistical information for traffic intensity and harvested energy. The problem is formulated as adapting BSs' on-off states, active resource blocks (e.g.~subcarriers) as well as power allocation to minimize the average grid power consumption in a given time period while satisfying the users' quality of service (blocking probability) requirements. It is transformed into an unconstrained optimization problem to minimize a weighted sum of grid power consumption and blocking probability. A \emph{two-stage dynamic programming (DP)} algorithm is then proposed to solve this optimization problem, by which the BSs' on-off states are optimized in the first stage, and the active BS's resource blocks are allocated iteratively in the second stage. Compared with the optimal joint BSs' on-off states and active resource blocks allocation algorithm, the proposed algorithm greatly reduces the computational complexity, while at the same time achieves close to the optimal energy saving performance.
\end{abstract}


\section{Introduction}
Exploiting renewable energy (e.g.~solar energy, wind energy and so on) from the surrounding environment to support wireless transmission data transmission, known as \emph{energy harvesting} technology, can support the operation of battery powered devices. Intelligently adapting the resource allocation of base stations (BSs) with energy harvesting equipment is a candidate solution to reduce the network energy consumption \cite{David:2010}. However, due to the limited availability of harvested energy as well as the uncertainty about the timing and the quantity of energy collected, there is a tradeoff between the quality of service (QoS) and the available power budget. Specifically, increasing the active wireless resource enhances the system capacity and users' service experience, but at the same time increases the probability of energy depletion, which will ultimately degrade users' QoS since the wireless resources may have to be powered down. Hence, in energy harvesting systems, wireless resource allocation should be optimized jointly considering the traffic profile, the users' QoS requirement, and the renewable energy statistics.

Resource allocation for energy harvesting systems has be extensively studied recently. J.~Yang \emph{et.~al.}~analyzed the offline optimal power allocation policy in a non-fading channel \cite{Jing:2012}. In the fading channel, the optimal power allocation is interpreted as the \emph{directional water-filling} policy \cite{Ozel:2011}. The offline analysis is extended to broadcast channel \cite{Jing:2012B}, multiple access channel \cite{Jing:2012M} and MIMO channel \cite{Maria:2013}. However, in practice, the energy arrival profile can not be known in advance due to uncertainty concerning the energy source. Consequently, the offline optimal policy is not applicable in real systems.

A practical way is to optimize the resource allocation using statistical information for harvested energy, for instance, the average arrival rate or the statistical distribution. Ref.~\cite{Gatz:2010} considers a cross-layer resource allocation problem to maximize the total system utility using a Markov decision process (MDP) approach \cite{Dimi:2005}. The packet dropping and blocking probabilities are analyzed with different sleep and wake-up strategies using queuing theory in sensor/mesh networks with solar power \cite{Niyato:2007}. In \cite{Bhargav:2009}, it is shown that the wireless link performance is strongly influenced by the renewable energy profile, and parameter adaptation is considered to improve the performance. The closed-form maximum stable throughput is studied and derived in cognitive radio networks \cite{Niko:2012} and cooperative networks \cite{Kidis:2012}, respectively. Nevertheless, most of existing work focuses on link level analysis, while the problem of how to efficiently allocate wireless resources according to the network traffic profile and the harvested energy profile from network point of view still remains open.

Based on the measured data, the statistics of the network traffic profile \cite{Will:2009, Auer:2011} and the harvested energy profile \cite{Maria:2011, Marco:2013} have been studied. In this paper, we make use of the statistical information for traffic intensity and harvested energy to study the wireless resource allocation problem in cellular networks. A mixed power supply from both renewable energy sources and power grid is adopted, which is considered as a candidate solution to minimize the energy consumption while at the same time guaranteeing users' QoS. Specifically, the reliable grid power guarantees that the service requirement is satisfied, while effective renewable energy allocation policy reduces the grid power consumption. In the literature, power allocation \cite{Jie:2013}, coordinated MIMO \cite{Cui:2012} and network planning \cite{Meng:2013} has been studied in the mixed power scenario. Different from the existing work, we aim to optimize the network operation according to the long-term network information with mixed power supply. Specifically, we consider the grid power minimization problem with users' QoS constraints in a downlink cellular network by adjusting BSs' on-off states and resource blocks allocation, where the BSs are equipped with energy harvesting devices. The preliminary results in single-cell case have been presented in \cite{Jie:2012}. This paper extensively studies the problem in multi-cell case. The main contents and contributions are listed as follows:

\begin{itemize}
 \item[$\bullet$] We formulate the problem of average grid power minimization taking into account the users' QoS (weighted blocking probability) constraints for a pre-defined time period (e.g.~24 hours), using knowledge of the traffic load profile and the energy harvesting statistics. The blocking probability is analyzed based on Erlang's approximation method \cite{Karray:2010} jointly considering the BSs' on-off states and the harvested energy profile.
 \item[$\bullet$] The grid power minimization problem is transformed into an unconstrained problem of minimizing a weighted combination of grid power consumption and blocking probability, which can be solved by a dynamic programming (DP) approach \cite{Dimi:2005}.
 \item[$\bullet$] A \emph{two-stage DP algorithm}, which determines the BSs' on-off state in the first stage, and then optimizes per-BS resource allocation in the second stage, is proposed to reduce the computational complexity. The performance of the proposed algorithm is evaluated by simulations and compared with the optimal DP algorithm and some heuristic algorithms.

\end{itemize}

The rest of the paper is organized as follows. Section \ref{sec:model} introduces the system model. The blocking probability is defined and analyzed in Section \ref{sec:blocking}. In Section \ref{sec:prob}, we study the average grid power minimization problem with a weighted blocking probability constraint. Numerical results are presented in Section \ref{sec:simu}. Finally, Section \ref{sec:concl} concludes the paper.

\section{System Model} \label{sec:model}
We consider a wireless cellular system with a total of $B$ BSs denoted as $\mathcal{B} = \{1,2,\ldots,B\}$, each of which is powered jointly by an energy harvesting device and the power grid. The operational time line (e.g. a period of 24 hours) is divided into $T$ time slots. The power model, the traffic model and the channel model are detailed as follows.

\subsection{Power Consumption Model} \label{sec:pmodel}
In slot $t$, the average harvested power of BS $b$ is denoted by $P_{H,t}^{(b)}$, and the grid power is $P_{G,t}^{(b)}$. Assume the harvested energy is stored in an infinite capacity battery. The assumption is reasonable as the harvested energy is generally not sufficient for reliable network operation. Hence, in a real system, even though the battery capacity is finite, there is only a very low probability of battery overflow. The BS energy consumption in \emph{active mode} is modeled as a constant power term plus a radio frequency (RF) related power \cite{Auer:2011}, which is
\begin{equation}
P_{BS,t}^{(b)} = P_0 + \Delta_PP_{RF,t}^{(b)}, \label{eq:PBS}
\end{equation}
where $P_0$ is the constant power including the baseband processor, the converter, the cooling system, and etc., $\Delta_P$ is the inverse of power amplifier efficiency factor, and $P_{RF,t}^{(b)}$ is the total RF transmit power.

Assume the total wireless bandwidth $W_0$ is divided into $N$ orthogonal subcarriers. The network will decide which BSs are powered on and how many subcarriers of these BSs are activated. The RF power is a linear function of the number of active subcarriers $n_{t}^{(b)}$, i.e.,
\begin{equation}
P_{RF,t}^{(b)} =\frac{n_{t}^{(b)}}{N}P_{T}, \quad n_{t}^{(b)} \le N, \label{eq:Pout}
\end{equation}
where $P_{T}$ is the constant transmit power level. Substituting $P_{RF,t}^{(b)}$ in (\ref{eq:PBS}) with (\ref{eq:Pout}), we get
\begin{equation}
P_{BS,t}^{(b)} = P_0 + \frac{n_{t}^{(b)}}{N}\Delta_PP_{T}. \label{eq:pfreq}
\end{equation}

In order to balance the performance among different time slots, the harvested energy may be reserved in the energy battery for future use by reducing the number of active subcarriers or by switching to \emph{sleep mode}. In this paper, two types of sleep modes are considered. The first one is deep {sleep mode}, in which a BS is completely turned off for a time slot. In this sleep mode, the BS power consumption is negligible and the users in the sleeping cell are served by the neighboring BSs. The second one is \emph{opportunistic sleep mode}. An active BS will turn to opportunistic sleep mode for a time ratio $\varphi_t^{(b)} \in [0,1)$ of the active period due to the lack of available energy input. It can be realized by time domain BS sleep \cite{Rui:2011} where some subframes are turned off. We assume the power consumption in opportunistic sleep mode is $P_S$. Denote $S_t^{(b)}$ as the state of BS $b$ at time $t$, which equals 1 if it is in active mode, and equals 0 otherwise. We summarise the BS state and power consumption model as follows:
\begin{equation}
P_{BS,t}^{(b)} = \left\{ \begin{array}{ll} P_0 + \frac{n_{t}^{(b)}}{N}\Delta_PP_{T}, & \textrm{if~} S_t^{(b)} = 1,\\
P_S, & \textrm{if~} S_t^{(b)} = 1 \textrm{~with opportunistic sleep,}\\
0, & \textrm{if~} S_t^{(b)} = 0.\end{array}\right.
\end{equation}
In reality, a BS in sleep mode still consumes a certain amount of power so that it can be reactivated. However, the power to reactivate a BS is negligible compared with the power consumption in active mode. Hence, the sleep mode power consumption is approximated as zero.

\subsection{Traffic Model}
In the following part of this section, we ignore the time index $t$ for simplicity. The users are sorted into groups according to their rate requirements and locations \cite{Balaji:2011}. Assume there are $K$ classes of users, each of which shares the same data rate requirement $R_k, k = 1, \ldots, K$. The network is further divided into $M$ disjoint regions, whose areas are denoted by $A_m, m = 1, \ldots, M$. In each region $m$, the users from class $k$ are uniformly distributed and randomly arrive according to a Poisson distribution with arrival rate $\lambda_{mk}$. Correspondingly, the service rate is denoted by $\mu_{mk}$. Hence, the traffic intensity of user class $k$ in area $m$ is calculated by
\begin{equation}
\rho_{mk} = \frac{\lambda_{mk}}{\mu_{mk}}.
\end{equation}

All the traffic in each area is served by the BS with largest signal strength. Once the BSs' active/sleep states are fixed, the serving BS for each user in each area is decided. The resource allocation of each BS follows the processor-sharing queueing model \cite{Klein:1976}, i.e., the active subcarriers are allocated to each user to meet its data rate requirement. Hence, a newly arrived user will be blocked if the available subcarriers are not sufficient to satisfy its rate requirement. Intuitively, when the network traffic load is high, more BSs and subcarriers should be active so that each BS takes care of a smaller area to guarantee the QoS. Otherwise, fewer BSs and subcarriers are required to be active, and hence the power consumption can be reduced.

\subsection{Channel Model}
We assume small-scale fast fading will average out as we consider the long time-scale performance, and large-scale shadowing will average out for sufficient cell realisation. Hence, we mainly focus on pathloss effects. The received SINR of user $u$ in the coverage of active BS $b$ is
\begin{equation}
\textrm{SINR}_u = \frac{P_T\beta (d_{u}^{(b)})^{-\alpha}}{\sigma^2 + \sum_{b':S^{(b')}=1,b'\neq b}\frac{n^{(b')}}{N} P_T\beta (d_{u}^{(b')})^{-\alpha}},
\end{equation}
where $\beta$ is the pathloss constant, $\alpha$ is the pathloss exponent, $d_{u}^{(b)}$ is the distance between BS $b$ and user $u$, and $\sigma^2$ is the noise power. Notice that we assume the interference is averaged over the whole bandwidth. That is, the perceived interference power is scaled by the ratio of active subcarriers $\frac{n^{(b')}}{N}$.
Then the maximum achievable transmission rate is
\begin{equation}
r_u = \frac{n^{(b)}W_0}{N}\log_2(1+\textrm{SINR}_u). \label{eq:rate}
\end{equation}

In the next section, we select the blocking probability as the QoS metric, and study the relationship between the blocking probability and the energy consumption.

\section{Blocking Probability Analysis} \label{sec:blocking}
The blocking probability is defined as the probability that a newly arrived user is blocked due to the lack of required radio resources. In energy harvesting systems, a blocking event may be caused by two factors. The first one is the high traffic load which results in that the required subcarriers are not available. We call the blocking due to the high traffic load as the \emph{service blocking probability}. The second one is the BS's opportunistic sleep mode in which a newly arrived user will be blocked. Hence, the blocking probability caused by opportunistic sleep is equal to the sleep ratio $\varphi_t^{(b)}$. We first analyze the service blocking probability, and then calculate the overall blocking probability.

\subsection{Service Blocking Probability}

Denote the instantaneous set of users of class $k$ in area $m$ by $\mathcal{U}_{mk}$, which are uniformly distributed in area $m$, and the user number by $U_{mk} = |\mathcal{U}_{mk}|$. We calculate the normalized bandwidth requirement of user $u$ of class $k$ in area $m$ by
\begin{equation}
\Phi_{mk}(u) = \frac{R_k}{r_{u}}.
\end{equation}
As each BS has a limited available bandwidth, the admission condition is that the total normalized bandwidth requirement denoted by $z_b$, should not exceed 1, i.e.,
\begin{equation}
z_b = \sum_{m\in \mathcal{M}^{(b)}}\sum_{k=1}^K \sum_{u \in \mathcal{U}_{mk}} \Phi_{mk}(u) < 1, \label{eq:admcond}
\end{equation}
where BS $b$ is assumed to be always active ($S^{(b)}=1, \varphi^{(b)} = 0$). Hence, the service blocking probability can be expressed as
\begin{align}
{p}_{\mathrm{sv},mk} = &\mathrm{Pr}(z_b < 1, z_b + \Phi_{mk}(u) \ge 1) \label{eq:pblkdef}\\
=&\mathrm{Pr}(1- \Phi_{mk}(u) \le z_b < 1), \label{eq:pblkori}
\end{align}
where (\ref{eq:pblkdef}) means that the total normalized bandwidth does not exceed 1 until a user $u$ of class $k$ arrives in area $m$. Calculation of the blocking probability (\ref{eq:pblkori}) requires the integration of the probability over all the possible quantities and locations of users served by BS $b$, for which it is difficult to find analytical expressions. We make use of the \emph{Erlang's approximation} method proposed in \cite{Karray:2010} and extend to our multi-class multi-area scenario. The basic idea of Erlang's approximation method is to average the users' bandwidth requirements over all the possible positions. Assuming that all the users in the area have the same bandwidth requirement, the blocking probability can be calculated by Erlang's formula \cite{Klein:1976}. Specifically, the average normalized bandwidth requirement of class $k$ users in area $m$ is
\begin{equation}
\bar{\Phi}_{mk} = \int_{A_m}\frac{R_k}{r_{u(a)}A_m}\mathrm{d}a.
\end{equation}
where $r_{u(a)}$ is the achievable data rate of user $u$ at position $a$, which is expressed as (\ref{eq:rate}). Hence, the admission condition (\ref{eq:admcond}) is changed to
\begin{equation}
\sum_{m\in \mathcal{M}^{(b)}}\sum_{k=1}^K U_{mk}\bar{\Phi}_{mk} < 1, \label{eq:admcondapprx}
\end{equation}
where $U_{mk}$ is the number of active users of class $k$ in area $m$. At the same time, the blocking probability of a user of class $k$ in area $m$ is modified as
\begin{equation}
{p}_{\mathrm{sv},mk} =\mathrm{Pr}(1- \bar{\Phi}_{mk} \le \sum_{m'\in \mathcal{M}^{(b)}}\sum_{k'=1}^K U_{m'k'} \bar{\Phi}_{m'k'} < 1). \label{eq:pblkapprx}
\end{equation}

According to the queueing theory \cite{Klein:1976}, the stationary probability of active user state $\bm{U}^{(b)} = \{U_{mk}\}_{m\in \mathcal{M}^{(b)},k=1,\ldots,K}$ associating to BS $b$ is
\begin{equation}
\pi^{(b)}(\bm{U}^{(b)}) = \prod_{m\in \mathcal{M}^{(b)}}\prod_{k=1}^K\frac{\rho_{mk}^{U_{mk}}}{U_{mk}!}\left( \sum_{\bm{U}^{(b)} \in \mathcal{U}^{(b)}} \prod_{m\in \mathcal{M}^{(b)}}\prod_{k=1}^K\frac{\rho_{mk}^{U_{mk}}}{U_{mk}!} \right)^{-1},
\end{equation}
where $\mathcal{U}^{(b)} = \{\bm{U}^{(b)}|\sum_{m\in \mathcal{M}^{(b)}}\sum_{k=1}^K U_{mk}\bar{\Phi}_{mk} < 1\}$ is the set of all possible active user states which satisfy the bandwidth constraint (\ref{eq:admcondapprx}). As a consequence, the blocking probability can be calculated as
\begin{equation}
{p}_{\mathrm{sv},mk} = \sum_{\bm{U}^{(b)} \in \bar{\mathcal{U}}^{(b)}_{mk}} \pi^{(b)}(\bm{U}^{(b)}), \quad m\in \mathcal{M}^{(b)}, \label{eq:pservice}
\end{equation}
where $\bar{\mathcal{U}}^{(b)}_{mk} = \{\bm{U}^{(b)}: 1-\Phi_{mk} \le \sum_{m'\in \mathcal{M}^{(b)}}\sum_{k'=1}^K U_{m'k'} \bar{\Phi}_{m'k'} \le 1\}$
is the set of active user states where the newly arrived user of class $k$ in area $m$ is blocked. In addition, the probability that a newly arrived user in the coverage of BS $b$ is blocked is
\begin{equation}
{p}_{\mathrm{sv}}^{(b)} = \frac{\sum_{m \in \mathcal{M}^{(b)}} \sum_{k=1}^K {p}_{\mathrm{sv},mk}\rho_{mk}}{\sum_{m \in \mathcal{M}^{(b)}} \sum_{k=1}^K \rho_{mk}} \label{eq:psvb}
\end{equation}


Notice that the service blocking probability can be tuned by adapting the BSs' working states $S^{(b)}, b = 1, \ldots, B$ and the number of active subcarriers in the active BSs $n^{(b)}, b \in \{b: S^{(b)} = 1\}$.

\subsection{Relation between $P_G^{(b)}$ and $\varphi^{(b)}$}
Recall that the BS in active mode can turn to opportunistic sleep mode with time ratio $\varphi^{(b)}$. Denote $P_{\mathrm{In}}^{(b)}$ as the total input power for transmission. According to the balance between power input and consumption, we have
\begin{equation}
P_{\mathrm{In}}^{(b)} = (1-\varphi^{(b)})P_{BS}^{(b)} + \varphi^{(b)}P_S, \quad P_{\mathrm{In}}^{(b)} \le P_{BS}^{(b)}
\end{equation}
Then the relation between the opportunistic sleep time ratio and the input power is
\begin{equation}
\varphi^{(b)} = \frac{P_{BS}^{(b)}-P_{\mathrm{In}}^{(b)}}{P_{BS}^{(b)}-P_S}. \quad P_{\mathrm{In}}^{(b)} \le P_{BS}^{(b)}
\end{equation}

We now discuss the relationship between the opportunistic sleep time ratio and the grid power consumption according to the available harvested power.

\subsubsection{Case 1} If the harvested energy is sufficient for the required input power, i.e., $P_{C}^{(b)} + P_{H}^{(b)} \ge P_{\mathrm{In}}^{(b)},$ where $P_C^{(b)}$ is the power supply from the battery. Then we have
\begin{equation}
P_G^{(b)} = 0, \quad \forall \; 0 \le \varphi^{(b)} \le 1,
\end{equation}
which means that the grid power input is not needed.

\subsubsection{Case 2} On the other hand, for the case where $P_{C}^{(b)} + P_{H}^{(b)} < P_{\mathrm{In}}^{(b)},$ then grid power is needed. And the opportunistic sleep time ratio can be expressed in terms of the grid power $P_G^{(b)}$ as
\begin{equation}
\varphi^{(b)} = \frac{P_{BS}^{(b)}-(P_{C}^{(b)} + P_{H}^{(b)} + P_G^{(b)})}{P_{BS}^{(b)}-P_S}, \quad P_G^{(b)} \le P_{BS}^{(b)} - (P_{C}^{(b)} + P_{H}^{(b)}).
\end{equation}

\subsection{Overall Blocking Probability}
In an energy harvesting system, if the BS is in opportunistic sleep mode due to the lack of energy, a newly arrived user of class $k$ in area $m$ served by BS $b$ will be blocked with probability 1. Otherwise, it will be blocked with probability ${p}_{\mathrm{sv},mk}$. As a consequence, the overall blocking probability can be calculated as
\begin{align}
p_{\mathrm{blk},mk} &= \varphi^{(b)} + (1-\varphi^{(b)}) {p}_{\mathrm{sv},mk}\nonumber\\
&= 1-(1-{p}_{\mathrm{sv},mk})(1-\varphi^{(b)}).
\end{align}
If we focus on the blocking probability for BS $b$, then we have
\begin{align}
{p}_{\mathrm{blk}}^{(b)} &= \frac{\sum_{m \in \mathcal{M}^{(b)}} \sum_{k=1}^K {p}_{\mathrm{blk},mk}\rho_{mk}}{\sum_{m \in \mathcal{M}^{(b)}} \sum_{k=1}^K \rho_{mk}} \nonumber\\
&= 1-(1-{p}_{\mathrm{sv}}^{(b)})(1-\varphi^{(b)}), \label{eq:pblk}
\end{align}
where ${p}_{\mathrm{sv}}^{(b)}$ is expressed as (\ref{eq:psvb}). Notice that if BS $b$ is in sleep mode ($S^{(b)} = 0$), the users in the sleeping cell must be associated with the other active BSs. Hence, no blocking events are counted for this sleeping BS.

\section{Power Grid Energy Minimization} \label{sec:prob}

\subsection{Problem Formulation}
In this section, we formulate the grid power minimization problem. The traffic intensity in time slot $t$ for the $K$ classes of users and $M$ regions is denoted by an $M\times K$ matrix $\bm{\rho}_t = \{\rho_{mk,t}\}_{m = 1, \ldots, M, k=1,\ldots,K}$. The energy harvesting power is denoted by a $1\times B$ vector $\bm{P}_{\bm{H},t} = [P_{H,t}^{(1)}, P_{H,t}^{(2)}, \ldots, P_{H,t}^{(B)}]$. The values of $\bm{\rho}_t$ and $\bm{P}_{\bm{H},t}$ are assumed to be constant for each slot $t$, but can vary among slots. By adjusting the BSs' on-off states $\bm{S}_t = [S_t^{(1)}, S_t^{(2)}, \ldots , S_t^{(B)}]$, the number of active subcarriers of active BSs $\bm{n}_t = [n_t^{(1)}, n_t^{(2)}, \ldots, n_t^{(B)}]$, and the opportunistic sleep time ratio $\bm{\varphi}_t = [\varphi_t^{(1)}, \varphi_t^{(2)}, \ldots, \varphi_t^{(B)}]$, we can adapt the grid power input as well as the total power usage in all the slots $t = 1, 2, \ldots, T$.

The following optimization problem is considered: given the traffic profile $\bm{\rho}_1, \bm{\rho}_2, \ldots, \bm{\rho}_T$ and the renewable energy profile $\bm{P}_{\bm{H},1}, \bm{P}_{\bm{H},2}, \ldots, \bm{P}_{\bm{H},T}$, adjust the BSs' working state $\bm{S}_1, \bm{S}_2, \ldots, \bm{S}_T$, the resource allocation $\bm{n}_1, \bm{n}_2, \ldots, \bm{n}_T$ and the sleep ratio $\bm{\varphi}_1, \bm{\varphi}_2, \ldots, \bm{\varphi}_T$ to minimize the average grid power consumption while satisfying the weighted blocking probability. Denote $\bm{S} = \{\bm{S}_1, \bm{S}_2, \ldots, \bm{S}_T\}, \bm{n} = \{\bm{n}_1, \bm{n}_2, \ldots, \bm{n}_T\}, \bm{\varphi} = \{\bm{\varphi}_1, \bm{\varphi}_2, \ldots, \bm{\varphi}_T\}$, then the problem can be formulated as
\begin{align}
\min\limits_{\bm{S}, \bm{n}, \bm{\varphi}} \;\; & \frac{\sum_{t=1}^TL_{t}\sum_{b=1}^B P_{G,t}^{(b)}}{\sum_{t=1}^T L_{t}} \label{prob:PGmin} \\
\mathrm{s.t.}\quad & \sum_{t=1}^T\sum_{b=1}^B \omega_{t}^{(b)} p_{\mathrm{blk},t}^{(b)} \le p_{\mathrm{target}}, \label{constr:pout}
\end{align}
where $L_t$ denotes the length of slot $t$, the blocking probability $p_{\mathrm{blk},t}^{(b)}$ is expressed as (\ref{eq:pblk}), and the weighting factor $\omega_{t}^{(b)}$, which satisfies $\sum_{t=1}^{T}\sum_{b=1}^B \omega_{t}^{(b)} = 1$, reflects the system sensitivity to the blocking probability in each slot. The weighting factor allows for the case where users may require higher QoS at some particular times of the day. For instance, if higher QoS is required during peak load times (e.g.~day time) than low load times (e.g.~night time), we can set the weighting factor for the day time to be larger than that for night time. The influence of the weighting factor settings is studied in the simulations.

\subsection{Optimal DP Algorithm}

The optimal solution for the problem (\ref{prob:PGmin}) with the constraint (\ref{constr:pout}) can be found by exhaustive search through all possible policies. However, this approach is not practical due to its high complexity. The DP approach \cite{Dimi:2005}, which divides the whole problem into simple per-stage sub-problems, is a candidate approach to find the optimal policy. We consider the following unconstrained optimization problem with a weighted combination of the power consumption and the blocking probability
\begin{equation}
\min\limits_{\bm{S}, \bm{n}, \bm{\varphi}}  \frac{\sum_{t=1}^TL_{t}\sum_{b=1}^B P_{G,t}^{(b)}}{\sum_{t=1}^T L_{t}} + \beta \sum_{t=1}^T\sum_{b=1}^B \omega_{t}^{(b)} p_{\mathrm{blk},t}^{(b)}, \label{prob:jointmin}
\end{equation}
where the factor $\beta > 0$ plays the role of a Lagrangian multiplier and indicates the relative importance of the blocking probability with respect to the average grid power consumption. Denote the minimum objective value of problem (\ref{prob:jointmin}) for a given $\beta$ as $P_{G\mathrm{ave},\beta}^* + \beta p_{\mathrm{blk},\beta}^*$, where $P_{G\mathrm{ave},\beta}^*$ and $p_{\mathrm{blk},\beta}^*$ represent the average grid power and the weighted blocking probability, respectively. As the objective (\ref{prob:jointmin}) is minimized, $P_{G\mathrm{ave},\beta}^*$ must be the minimum average grid power to guarantee that the blocking probability is no more than $p_{\mathrm{blk},\beta}^*$. Hence, the solution for (\ref{prob:jointmin}) is also the one for (\ref{prob:PGmin}) where $p_{\mathrm{target}} = p_{\mathrm{blk},\beta}^*$.

Denote $P_{G\mathrm{ave}} (p_{\mathrm{blk}})$ as the minimum average grid power such that the blocking probability does not exceed $p_{\mathrm{blk}}$. Hence, we have $P_{G\mathrm{ave}}(p_{\mathrm{blk},\beta}^*) = P_{G\mathrm{ave},\beta}^*$. By adjusting the value of $\beta$ and solving the corresponding problem (\ref{prob:jointmin}), we can find a set of points for the function $P_{G\mathrm{ave}} (p_{\mathrm{blk}})$. By joining these points which indicate the minimum average grid power for a given target blocking probability, we get a lower bound curve of the grid power consumption for the target blocking probability. Any achievable pair of grid power consumption and blocking probability values must lie above this curve. Notice that it is not guaranteed that all the values of $P_{G\mathrm{ave}} (p_{\mathrm{blk}})$ can be found. Hence, for a given target blocking probability $p_{\mathrm{target}}$, if a corresponding point can be found by setting appropriate value of $\beta$, the optimal solution for the original problem (\ref{prob:PGmin}) with constraint (\ref{constr:pout}) is found. Otherwise, we can just get a suboptimal result by adopting the policy related to the point with the largest blocking probability less than $p_{\mathrm{target}}$.

The DP algorithm contains three key components: state, action and cost function. In the problem (\ref{prob:jointmin}), the state is the amount of energy $\bm{E}_{\bm{C},t} = [E_{C,t}^{(1)}, E_{C,t}^{(2)}, \ldots, E_{C,t}^{(B)}]$ in the battery at the beginning of slot $t$. For each BS $b$, $E_{C,t}^{(b)}$ evolves to slot $t+1$ as follows:
\begin{align}
E_{C,t+1}^{(b)} = E_{C,t}^{(b)}+L_{t}(P_{H,t}^{(b)} + P_{G,t}^{(b)}) - \left[\left(1-\varphi_t^{(b)}\right)L_{t}P_{BS,t}^{(b)} + \varphi_t^{(b)}L_{t}P_S\right],
\end{align}
where the energy consumption in slot $t$ can not exceed the energy available, i.e.,
\begin{align}
\left[\left(1-\varphi_t^{(b)}\right)L_{t}P_{BS,t}^{(b)} + \varphi_t^{(b)}L_{t}P_S\right] \le E_{C,t}^{(b)}+L_{t}(P_{H,t}^{(b)} + P_{G,t}^{(b)}),
\end{align}
and the power grid is not plugged in until the harvested energy is not enough:
\begin{equation}
P_{G,t}^{(b)} \le \max\left\{ \left(1-\varphi_t^{(b)}\right)L_{t}P_{BS,t}^{(b)} + \varphi_t^{(b)}L_{t}P_S - E_{C,t}^{(b)} - L_{t}P_{H,t}^{(b)} \right\}.
\end{equation}
The actions are the BSs' working state $\bm{S}_t$, the number of active subcarriers $\bm{n}_t$, and the sleep ratio $\bm{\varphi}_t$. Notice that if $S_{t}^{(b)} = 0$, there is no active subcarrier ($n_{t}^{(b)} = 0$), and the BS keeps sleep during the slot $t$ ($\varphi_t^{(b)} = 1$). The per-stage cost is the weighted combination of the average grid power and the blocking probability, denoted as a function of the current action and state
\begin{equation}
c_{t}(\bm{S}_t, \bm{n}_t, \bm{\varphi}_t, \bm{E}_{C,t}) = \frac{L_{t}\sum_{b=1}^BP_{G,t}^{(b)}}{\sum_{t=1}^T L_{t}B} + \beta \sum_{b=1}^{B}\omega_{t}^{(b)} p_{\mathrm{blk},t}^{(b)}.
\end{equation}
The DP algorithm breaks the original problem down into sub-problems with respect to the stage, where the objective is to minimize the cost of each time slot plus that of the following slots. The per-slot sub-problems are solved recursively. The \emph{cost-to-go} function is defined recursively as
\begin{equation}
J_{t}(\bm{E}_{C,t}) = \left\{ \begin{array}{ll} \min\limits_{\bm{S}_t, \bm{n}_t, \bm{\varphi}_t}\! c_{t}(\bm{S}_t, \bm{n}_t,  \bm{\varphi}_t, \bm{E}_{C,t}), & t = T \\
\min\limits_{\bm{S}_t, \bm{n}_t, \bm{\varphi}_t}\! \left\{ c_{t}(\bm{S}_t, \bm{n}_t, \bm{\varphi}_t, \bm{E}_{C,t}) \!+\! J_{t+1}(\bm{E}_{C,t+1}) \right\}, & t < T\end{array}\right. \label{eq:costtogo}
\end{equation}
which denotes the minimum cost of the sub-problem with slot $t$ as its initial stage. Performing a backward induction of the cost-to-go functions (\ref{eq:costtogo}) from time slot $T$ to slot 1, we can obtain the minimum cost equal to $J_1(\bm{0})$.

Assume the number of examined sleep ratios $\varphi_t^{(b)}$ is $N_{\varphi}$. Then, the cardinality of the action space for each cost-to-go function is $(NN_{\varphi}+1)^B$. Note that the number of BS actions $(S_t^{(b)}, n_t^{(b)}, \varphi_t^{(b)})$ is $(NN_{\varphi}+1)$ instead of $2NN_{\varphi}$, as the BSs in sleep mode have only a single state. Hence, given the state in time slot $t$, the cardinality of the state space in slot $(t+1)$ is no more than $(NN_{\varphi}+1)^B$. That is, if the harvested energy of all BSs is enough for any resource allocation policy, each policy corresponds to a unique next-stage state. Otherwise, some policies result in the same state, so the state space is less than $(NN_{\varphi}+1)^B$.

Both the action space and the state space dimensions increase exponentially with the number of BSs $B$ in the network, which, due to the \emph{curse of dimensionality} \cite{Dimi:2005}, will result in an overwhelming computational complexity to find the optimal control policy if the network size is large. As a consequence, the proposed DP optimal algorithm is difficult to implement in practical systems, and low-complexity solutions are required. In the following, a two-stage optimization algorithm is proposed to reduce the size of state and action space.

\subsection{Two-stage DP Algorithm}
The basic idea of the two-stage optimization algorithm is to divide the action process into two steps. In the first stage, we assume that the number of active subcarriers in active BSs are always $N$, i.e., subcarrier allocation is not considered in this stage. In addition, the active BS sleep ratio $\varphi^{(b)}_t$ is assumed to be 0 for all the $b = 1, \ldots, B$, which means the required power is always available. As a result, the actions at this stage only consist of the BSs' working states $\bm{S}$. The optimization problem can be written as
\begin{equation}
\min\limits_{\bm{S}}  \frac{\sum_{t=1}^TL_{t}\sum_{b=1}^B P_{G,t}^{(b)}}{\sum_{t=1}^T L_{t}} + \beta \sum_{t=1}^T\sum_{b=1}^B \omega_{t}^{(b)} p_{\mathrm{blk},t}^{(b)} \Big|_{\bm{n}_t = \bm{N}, \bm{\varphi}_t = \bm{0}, \forall t}, \label{prob:1stage}
\end{equation}
where $\bm{N} = N\bm{S}_t$, and $\bm{0} = 1-\bm{S}_t$. Hence, the condition $\bm{n}_t = \bm{N}, \bm{\varphi}_t = \bm{0}$ means for any $b = 1, 2, \ldots, B$, if $S_t^{(b)} = 1$, the corresponding number of active subcarriers is $n_t^{(b)} = N$, and the sleep ratio is $\varphi_t^{(b)} = 0$, i.e., BS $b$ activates all the subcarriers for the whole time slot $t$. The cost-to-go function is
\begin{align}
& J_{t}(\bm{E}_{C,t})|_{\bm{n}_t = \bm{N}, \bm{\varphi}_t = \bm{0}} \nonumber\\
= & \left\{ \begin{array}{ll} \min\limits_{\bm{S}_t}\! c_{t}(\bm{S}_t, \bm{n}_t = \bm{N},  \bm{\varphi}_t = \bm{0}, \bm{E}_{C,t}), & t = T \\
\min\limits_{\bm{S}_t}\! \left\{ c_{t}(\bm{S}_t, \bm{n}_t = \bm{N},  \bm{\varphi}_t = \bm{0}, \bm{E}_{C,t}) \!+\! J_{t+1}(\bm{E}_{C,t+1})|_{\bm{n}_{t+1} = \bm{N}, \bm{\varphi}_{t+1} = \bm{0}} \right\}, & t < T\end{array}\right. \label{eq:costtogo1stage}
\end{align}

\emph{Remark 1}: The action space of each cost-to-go function in (\ref{eq:costtogo1stage}) is $2^B$, and given the state in time slot $t$, the maximum state space in time slot $(t+1)$ is reduced from $(NN_{\varphi}+1)^B$ to $2^B$.

The problem (\ref{prob:1stage}) can be solved by the standard DP algorithm with a much lower complexity compared to the original DP problem (\ref{prob:jointmin}). In the second stage, given the BSs' working state $\bm{S}^* = \{\bm{S}_1^*, \bm{S}_2^*, \ldots, \bm{S}_T^*\}$ obtained from the first stage, we adjust the number of active subcarriers and power allocation for each BS separately. Since the subcarrier adaptation changes the interference profile, the per-BS resource allocation correlates with one another. We propose an iterative resource allocation algorithm, which updates the per-BS resource allocation based on the allocation results of the other BSs, and then iterates the process until the resource allocation solution does not change between two consecutive iterations. The per-BS resource allocation optimization problem can be formulated as
\begin{equation}
\min\limits_{n_t^{(b)}, \varphi_t^{(b)}} \frac{\sum_{t=1}^TL_{t}\sum_{b=1}^B P_{G,t}^{(b)}}{\sum_{t=1}^T L_{t}} + \beta \sum_{t=1}^T\sum_{b=1}^B \omega_{t}^{(b)} p_{\mathrm{blk},t}^{(b)} \Big|_{\bm{S}^{*}, n_t^{(b')},\varphi_t^{(b')},\forall t, b'\neq b}. \label{prob:2stage}
\end{equation}
The problem can also be solved by the DP algorithm where the cost-to-go function is
\begin{align}
& J_{t}^{(b)}({E}_{C,t}^{(b)})|_{\bm{S}_t^{*}, n_t^{(b')},\varphi_t^{(b')}, b'\neq b} \nonumber\\
=& \left\{ \begin{array}{ll} \min\limits_{n_t^{(b)}, \varphi_t^{(b)}}\! c_{t}(\bm{S}_t = \bm{S}_t^{*}, \bm{n}_t^{(b)},  \bm{\varphi}_t^{(b)}, \bm{E}_{C,t}^{(b)})|_{ n_t^{(b')}, \varphi_t^{(b')}, b'\neq b}, & t = T \\
\min\limits_{n_t^{(b)}, \varphi_t^{(b)}}\! \left\{ c_{t}(\bm{S}_t = \bm{S}_t^{*}, \bm{n}_t^{(b)},  \bm{\varphi}_t^{(b)}, \bm{E}_{C,t}^{(b)})|_{ n_t^{(b')}, \varphi_t^{(b')}, b'\neq b} \!+\! J_{t+1}(\bm{E}_{C,t+1})|_{\bm{S}_t^{*}, n_{t+1}^{(b')}, \varphi_{t+1}^{(b')}, b'\neq b} \right\}, & t < T\end{array} \right. \label{eq:costtogo2stage}
\end{align}

\emph{Remark 2}: The action space of each cost-to-go function in (\ref{eq:costtogo2stage}) is either $NN_{\varphi}$ ($S_t^{(b)*} = 1$) or 1 ($S_t^{(b)*}=0$), and given the state in time slot $t$, the maximum state space in time slot $(t+1)$ is no more than $NN_{\varphi}$.

\emph{Remark 3}: In summary, using the two-stage optimization algorithm, the action space of each time slot optimization is reduced from $(NN_{\varphi}+1)^B$ to $2^BBNN_{\varphi}$. Accordingly, given the state in time slot $t$, the maximum state space in time slot $(t+1)$ is reduced from $(NN_{\varphi}+1)^B$ to $2^BBNN_{\varphi}$.

The two-stage optimization algorithm is summarized as Algorithm \ref{alg:twostage}.

\begin{algorithm}[h] \caption{Two-stage DP Optimization}
\label{alg:twostage}
\begin{algorithmic}

\STATE \textbf{The 1st stage}:

\STATE Solve the problem (\ref{prob:1stage}) to find $\bm{S}^*$.

\STATE \textbf{The 2nd stage}:

\STATE Set $\bm{n}_t = \bm{N}, \bm{\varphi}_t = \bm{0}, \bm{{n}'}_t \neq \bm{n}_t, \bm{{\varphi}'}_t \neq \bm{\varphi}_t, t = 1, \ldots, T$

\WHILE{$\bm{n}_t \neq \bm{{n}'}_t$ or $\bm{\varphi}_t \neq \bm{{\varphi}'}_t$ for some $t = 1, \ldots, T$}

\STATE Set $\bm{{n}'}_t = \bm{n}_t, \bm{{\varphi}'}_t = \bm{\varphi}_t, t = 1, \ldots, T$.

\FOR {$b = 1$ to $N$}

\STATE Set $\bm{n}^{(b)} = \{n_1^{(b)}, n_2^{(b)}, \ldots, n_T^{(b)}\}, \bm{\varphi}^{(b)} = \{\varphi_1^{(b)}, \varphi_2^{(b)}, \ldots, \varphi_T^{(b)}\}$

\STATE Find $\bm{n}^{(b)*}, \bm{\varphi}^{(b)*}$ which solve the problem (\ref{prob:2stage}) by fixing $\bm{n}^{(b')}$ and $\bm{\varphi}^{(b')}, b' \neq b$.

\STATE Update $\bm{n}_t,\bm{\varphi}_t, t = 1, \ldots, T$ by setting $\bm{n}^{(b)} = \bm{n}^{(b)*}, \bm{\varphi}^{(b)} = \bm{\varphi}^{(b)*}$.

\ENDFOR

\ENDWHILE

\end{algorithmic}
\end{algorithm}

\subsection{Heuristic Algorithms}
Motivated by the two-stage DP algorithm where the BSs' on-off states are determined in the first stage, and the per-BS resource allocation is determined in the second stage, we propose some low-complex heuristic algorithms for comparison, which also operates in two-stage manner. Specifically, the BSs' on-off states can be adjusted by the following algorithms:
\begin{itemize}
 \item[$\bullet$] \emph{Non-sleep policy}. In this policy, all the BSs are active in each slot. It is the policy used in the traditional cellular network, which can be viewed as a baseline strategy.
 \item[$\bullet$] \emph{Threshold-based sleep policy}. In this policy, the number of active BSs are decided by the network traffic intensity. Basically, there is a minimum number of active BSs required, $B_{\mathrm{min}}$, to guarantee the network coverage. We define a set of thresholds $\theta_0 (= 0), \theta_1, \ldots, \theta_Q, Q \le B-B_{\mathrm{min}}$. The BSs' on-off pattern is pre-defined for each $\theta_i, 1 \le i \le Q$. If the integrated network traffic intensity satisfies $\theta_{i-1} < \sum\limits_m \sum\limits_k \rho_{mk} \le \theta_{i}$, the corresponding BSs on-off pattern is selected.
\end{itemize}

Once the BSs' on-off states are decided, the number of active subcarriers and the opportunistic sleep ratio are tuned in each BS individually based on the algorithms listed below:
\begin{itemize}
 \item[$\bullet$] \emph{Maximum resource block utilization}. In this policy, all the blocks are activated for transmission, i.e., $n_{t}^{(b)} = N$ for all $t$. It can be considered as a baseline case.
 \item[$\bullet$] \emph{Traffic-aware resource block utilization}. Based on the intuition that higher traffic intensity requires more wireless resources, we propose a policy where the number of activated subcarriers is set proportional to the traffic intensity, i.e.,
     \begin{equation}
     n_{t}^{(b)} = \min\{N, \lceil\eta_1\rho_{t}N\rceil\},\quad \eta_1 > 0
      \end{equation}
      where $\lceil x \rceil$ is the minimum integer no smaller than $x$.
 \item[$\bullet$] \emph{Joint traffic-energy-aware resource block utilization}. As the outages are caused not only by the lack of wireless resources, but also by the lack of power, the power budget should be taken into consideration. In this case, the number of active subcarriers is also proportional to the available power besides the traffic intensity:
     \begin{equation}
     n_t^{(b)} = \min\left\{N, \left\lceil\eta_2 \rho_{t} \frac{E_{B,t}^{(b)} + E_{G,t}^{(b)}+ L_{t}P_{H,t}^{(b)} }{\sum_{k=t}^TL_k(P_0+\Delta_PP_T)}N \right\rceil \right\},
      \end{equation}
      where $\eta_2 \!>\! 0$, and the grid energy $E_{G,t}^{(b)}$ evolves as $E_{G,t+1}^{(b)} \!=\! \max\{0, E_{G,t}^{(b)}\! - \!L_tP_{G,t}^{(b)}\}$. Note that $P_0+\Delta_PP_T$ in the denominator is for normalization.
\end{itemize}

The sleep ratio $\varphi_t^{(b)}$ in all these policies is decided as follows. Given the average grid power $P_{G\mathrm{ave}}^{(b)}$, the grid energy budget is initialized as $E_{G,1}^{(b)}\! = \!\sum_{t=1}^TL_tP_{G\mathrm{ave}}^{(b)}$. We get
\begin{equation}
P_{G,t}^{(b)} \!=\! \min\left\{ \frac{E_{G,t}^{(b)}}{L_t}, \max\left\{0, P_0\! +\! \frac{n_t^{(b)}}{N}\Delta_PP_T \!- \!\frac{E_{B,t}^{(b)}}{L_t} \!-\! P_{H,t}^{(b)}\right\}\right\},
\end{equation}
i.e., the grid power is used to satisfy the power requirement as long as it is available.

Notice that given the parameters $\theta_i, \eta_1, \eta_2$, the heuristic algorithms only depend on the traffic and the energy conditions of current time slot. The complexity is much lower than the DP algorithm. However, the QoS performance is not guaranteed, which is shown in the simulation results in the next section.

\section{Numerical Simulations} \label{sec:simu}
We examine the performance of the proposed algorithms by numerical simulations. We adopt the energy consumption model of the macro BS from the EARTH project \cite{Auer:2011}, and the channel model from 3GPP LTE \cite{Sesia:2009}. In the macro-cell scenario, we have $P_0=712.2$W, $\Delta_P = 15.96$, the maximum transmit power $P_{\mathrm{max}} = 40$W, and the cell radius $R = 1000$m. The opportunistic sleep mode power is $P_S = 50$W. The bandwidth is set to $W_0 = 10$MHz and the number of sub-carriers is set to $N=600$. The path-loss is $\mathrm{PL}^{\textrm{dB}} = 34.5 + 35 \log_{10}(l)$, and the noise power density is $-174$dBm/Hz. We first study the relation between the QoS and the resource allocation in single cell scenario, and evaluate the proposed DP algorithm in this setup. Then the simulation is extended to sectorized multi-cell scenario to study the performance of the two-stage DP algorithm.

\begin{figure}
\centering
\includegraphics[width=3in]{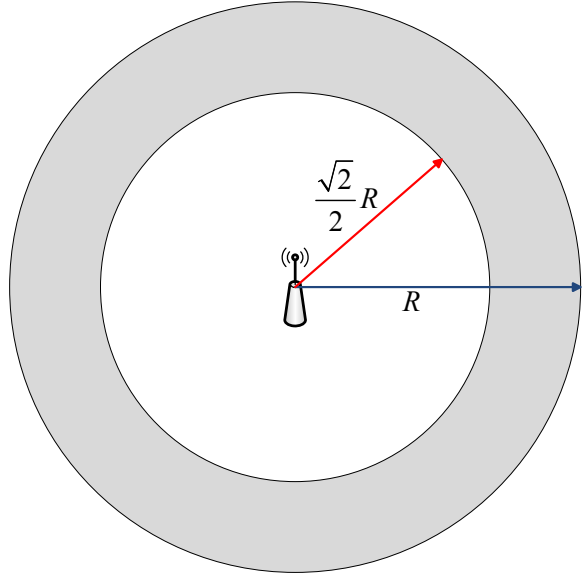}
\caption{Single-cell Erlang's approximation settings for $M=2$.} \label{fig:singlecell}
\end{figure}

\begin{figure}
\centering
\includegraphics[width=4.5in]{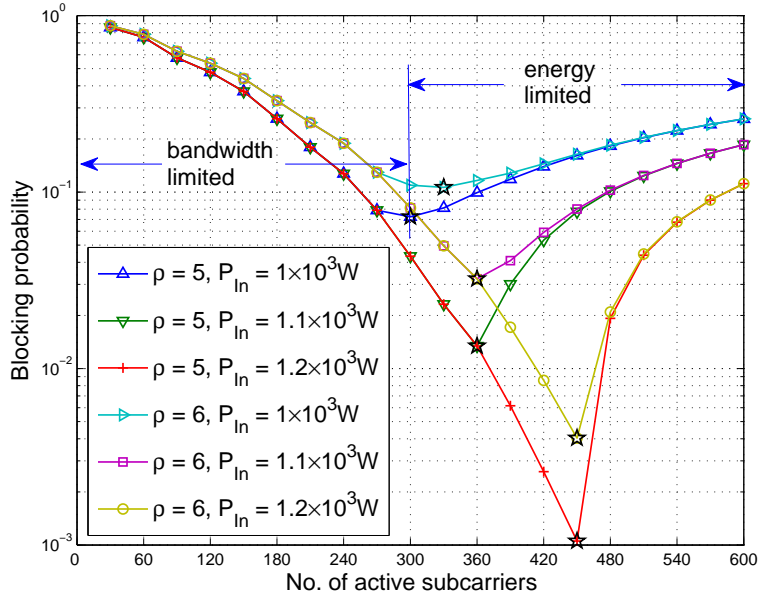}
\caption{Relationship between blocking probability and number of active subcarriers. $\rho$ is the traffic intensity, $P_{\mathrm{In}}$ is the total input power available. The ``star" is the minimum blocking probability for each parameter settings.} \label{fig:singleblkvsN}
\end{figure}

\subsection{Single-Cell Case}
For the single-cell case, the superscript $b$ is ignored for simplicity. We set the number of user classes as $K=1$. The circular cell area are divided into $M=2$ regions with equal areas, as shown in Fig.~\ref{fig:singlecell}. It is easy to find that the inner circular region is of radius of $\frac{\sqrt{2}}{2}R$. Accordingly, the user data requirement is $r_K = r_0 = 2$Mbps, the user service rate is $\mu_{1K} = \mu_{2K} = \mu = 1\mathrm{s}^{-1}$, and the arrival rate $\lambda_{1K} = \lambda_{2K} = \frac{\lambda}{2}$. The total traffic intensity is denoted by $\rho = \lambda/\mu$, and the total input power is denoted by $P_{\mathrm{In}}$, which includes harvested power, grid power and battery power. The relation between the number of active subcarriers and the blocking probability is depicted in Fig.~\ref{fig:singleblkvsN}. Take $\rho = 5, P_{\mathrm{In}} = 1\times 10^3W$ as an example. When the number of active subcarriers is less than 300, the blocking is mainly caused by the limited availability of subcarriers. Hence, the region where $n<300$ is called the \emph{bandwidth limited} region. On the contrary, if $n\ge300$, the available power is insufficient to enable the active subcarriers to be always on, which means $\varphi > 0$. Then the blocking is also caused by opportunistic sleep, which gradually becomes the main blocking factor. Correspondingly, the region where $n\ge300$ is called the \emph{energy limited} region. As a result, there is a minimum outage probability working point as shown by the star on each curve. In addition, if a certain blocking probability can be achieved in both bandwidth limited region and energy limited region, the policy in bandwidth limited region consumes the power less than $P_{\mathrm{In}}$, while that in energy limited region consumes all the available power $P_{\mathrm{In}}$. Hence, subcarrier adaptation according to the traffic requirement and available energy is more efficient than opportunistic sleeping. In the following simulation, we only consider subcarrier adaptation optimization, i.e., we set $\varphi = 0$ for all the conditions.

\begin{figure}
\centering
\includegraphics[width=4.5in]{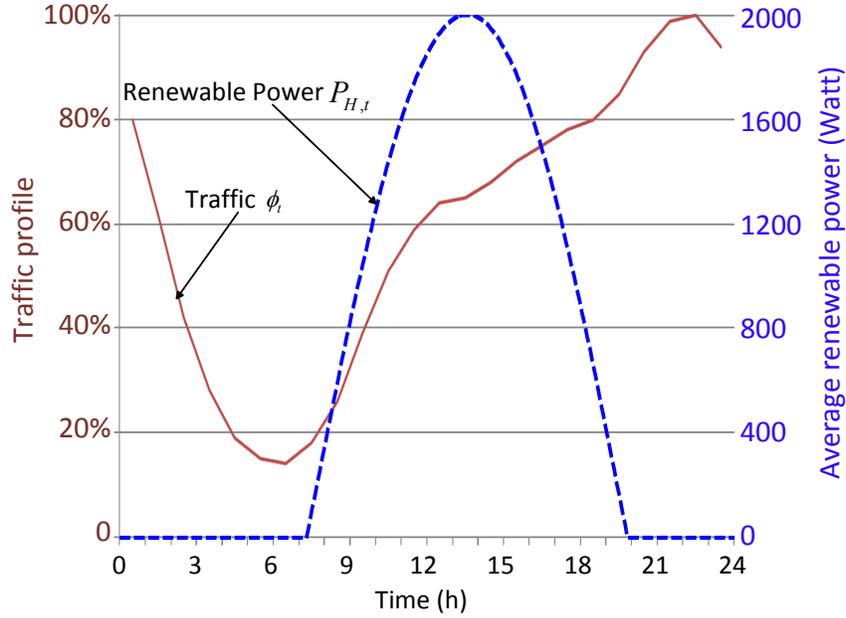}
\caption{Daily traffic (solid line) \cite{Auer:2011} and renewable energy profile (dashed line) \cite{Maria:2011}.} \label{fig:traffic}
\end{figure}

Then the performance of the DP algorithm to minimize the grid energy consumption is evaluated with a given traffic profile and energy arrival statistics for one day. We run the standard DP algorithm (\ref{eq:costtogo}) for the single-cell case as the computational complexity is affordable. The traffic profile and renewable energy harvesting profile are taken from \cite{Auer:2011} and \cite{Maria:2011}, respectively, as shown in Fig.~\ref{fig:traffic}. We set $T=24$
, and the length of each slot is $L_{t} = 1$ hour. The traffic profile is $\lambda_{t} = \phi_{t}\lambda_{\mathrm{max}}$, where the maximum traffic intensity $\lambda_{\mathrm{max}}=10\mathrm{s}^{-1}$ and $0< \phi_{t} \le 1$.

\begin{figure}
\centering
\includegraphics[width=4.5in]{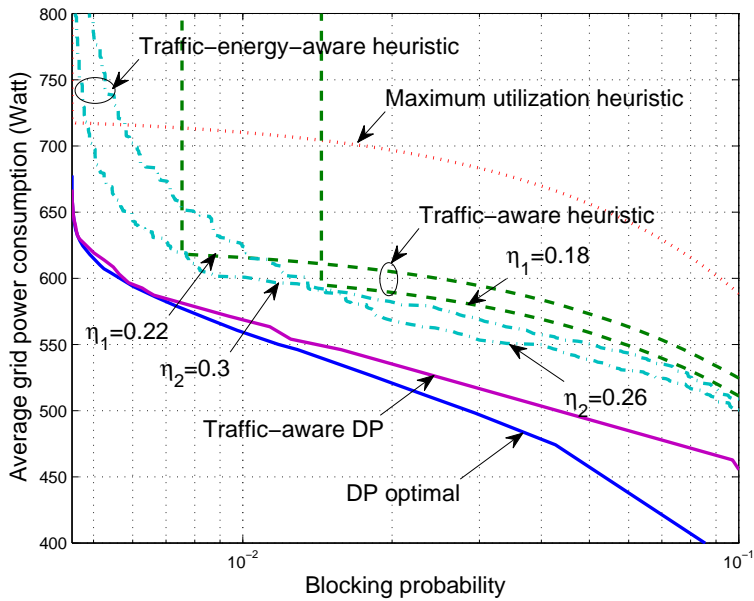}
\caption{Tradeoff curves between outage probability and grid energy consumption with different policies.} \label{fig:DPtradeoff}
\end{figure}

The tradeoff between average blocking probability ($\omega_{t} = 1/T$) and grid energy consumption for different policies is depicted in Fig.~\ref{fig:DPtradeoff}. Notice that the \emph{traffic-aware DP} algorithm firstly optimizes the resource allocation via DP approach assuming only grid power input, and then calculates the actual grid power consumption considering the renewable energy profile. It can be seen that the proposed DP based algorithm is the optimal solution, which verifies that the traffic variation and the energy profile should be jointly considered. The comparison between the traffic-aware heuristic algorithm and the traffic-energy-aware heuristic algorithm also confirms this. Specifically, the joint traffic-energy-aware policy performs better than the traffic-aware policy in almost all conditions by choosing proper $\eta_2$ ($\eta_2 = 0.26, 0.3$ for $\eta_1 = 0.18, 0.22$, respectively). In addition, by adjusting the value of $\eta_1$ and $\eta_2$, we obtain different curves.  For instance, the traffic-aware heuristic algorithm with a smaller value of $\eta_1$ (0.18) performs closer to the optimal than that with larger value (0.22) for the low grid power input regime ($<610$ Watt), and that with larger $\eta_1$ (0.22) is near optimal for the high grid energy input regime ($>610$ Watt).

\begin{figure}
\centering
\includegraphics[width=4.5in]{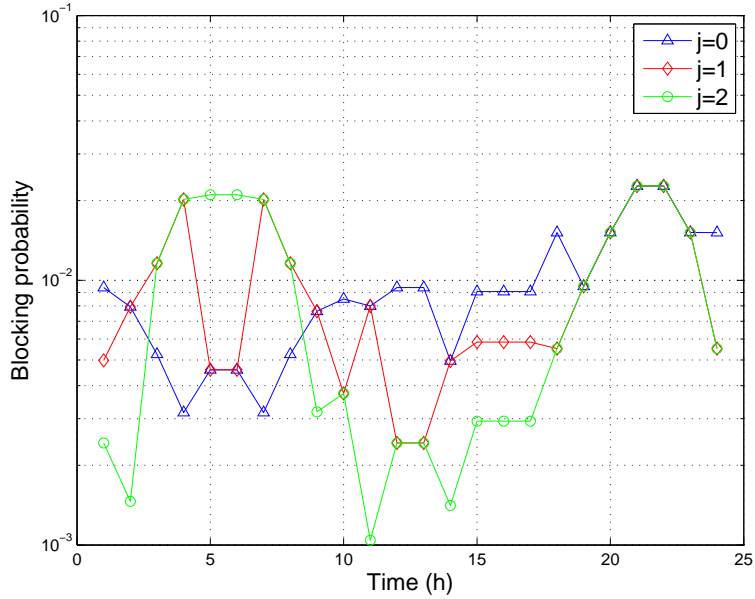}
\caption{Per-slot outage probability with different weighting factors $\omega_t = \phi_t^j/\sum_t \phi_t^j$. The average outage probability is identical as 1\%.} \label{fig:Optblocking}
\end{figure}

Fig.~\ref{fig:Optblocking} shows the per-slot blocking performance of the DP algorithm for the same average blocking probability target (1\%). In this simulation, we set $\omega_t = \phi_t^j/\sum_t \phi_t^j$, where $j = 0,1,2$. The exponent $j=0$ corresponds to the average blocking probability, and $j=1$ means traffic weighted blocking. As $\phi_t < 1$, larger $j$ implies a higher weight for the high traffic regime. It can be seen that by adjusting the weighting factor, we can obtain different blocking profiles. Specifically, the algorithm tends to increase the blocking probability in a low traffic load regime if the corresponding weighting factor is large ($j=2$).

\begin{figure}
\centering
\includegraphics[width=4in]{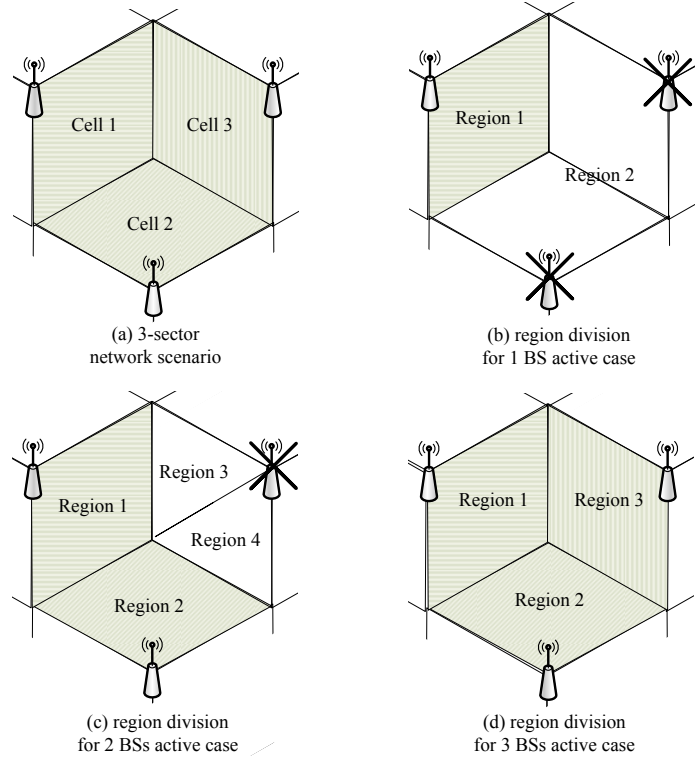}
\caption{Sectorized Multi-cell Erlang's approximation settings.} \label{fig:multicell}
\end{figure}

\subsection{3-Sector Case}
We now turn to multi-cell scenario. We consider the sectorized multi-cell setup as shown in Fig.~\ref{fig:multicell}(a), where each cite has 3 co-located BSs. In this setup, the dominant interference for a user in cell 1, 2, or 3 is from the other two cells. The interference from BSs at further locations can be considered as a low-power background noise. Hence, the optimization can be done in each 3-sector cluster individually.

The parameter settings are as follows. The regional division depends on the BSs' on-off state. Specifically, if only one BS is active, as shown in Fig.~\ref{fig:multicell}(b), the cluster is divided into $M=2$ regions. The first region is the original coverage of the active BS, and the second is that of the others. If two BSs are active (Fig.~\ref{fig:multicell}(c)), $M=4$. The coverage of the sleep BS is divided into 2 regions, which are served by the two active BSs, respectively. Finally, if all BSs are active (Fig.~\ref{fig:multicell}(d)), $M=3$, and each region is covered by its own BS. The service rate is assumed the same $\mu_K = 1\mathrm{s}^{-1}$. The traffic in the studied area follows the profile illustrated in Fig.~\ref{fig:traffic}, and each sector occupies part of it. Assume that the user arrival rate of sector $b$ in time slot $t$ is $\lambda_t^{(b)} = \psi^{(b)}\lambda_t$, where $0\le \psi^{(b)} \le 1, \sum_b \psi^{(b)} = 1$. We run the simulations for two setups: an asymmetric traffic distribution $\psi^{(1)}:\psi^{(2)}:\psi^{(3)} = 1:2:3$ and a symmetric distribution $\psi^{(1)}:\psi^{(2)}:\psi^{(3)} = 1:1:1$. The renewable energy profile is assumed identical among the three BSs as in reality, the renewable energy (ex. solar power) intensity will be almost the same in a cluster-sized region. In our simulation, the same energy profile depicted in Fig.~\ref{fig:traffic} is adopted for all the BSs.

\begin{figure}
\centering
\includegraphics[width=4.5in]{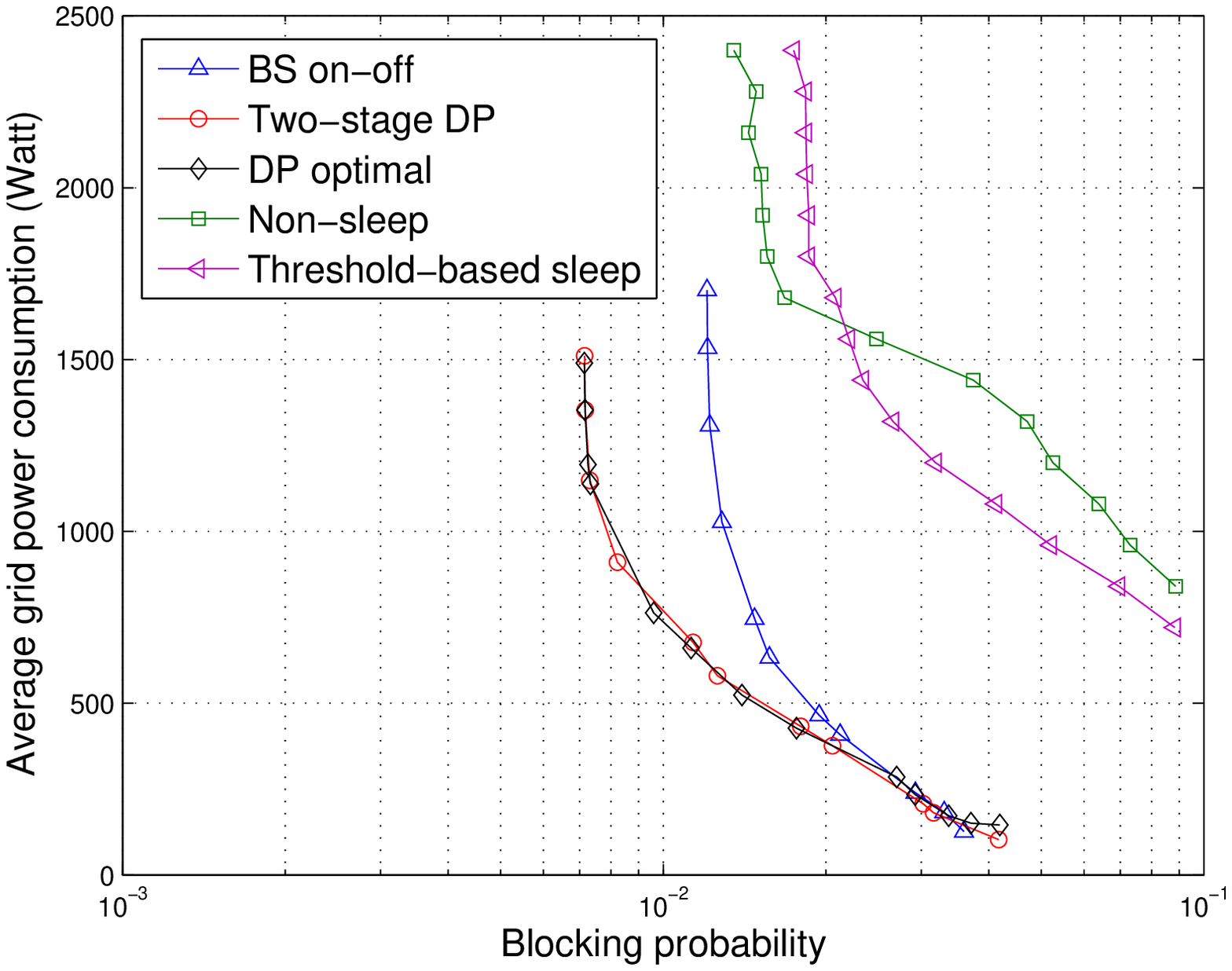}
\caption{Tradeoff curves between outage probability and grid energy consumption for 3-sector case. $K=1, R_1 = 2\textrm{Mbps}, \psi^{(1)}:\psi^{(2)}:\psi^{(3)} = 1:2:3$. The renewable energy profile as in Fig. \ref{fig:traffic} is the same for three BSs.} \label{fig:K1a}
\end{figure}

The tradeoff between blocking probability and gird energy consumption under an asymmetric traffic distribution for a single user class ($K=1$) is shown in Fig.~\ref{fig:K1a}. Notice that the \emph{BS on-off} algorithm only optimizes the BS on-off state assuming all subcarriers are active ($n_t^{(b)} = N$) so there is no opportunistic sleep ($\varphi_t^{(b)} = 0$). It is actually the first stage optimization in the two-stage algorithm. Also notice that for the heuristic non-sleep and threshold-based sleep algorithms, the joint traffic-energy aware adaptation algorithm is used in the second stage as it is better than the other heuristic algorithms. In the threshold-based sleep algorithm, we set two thresholds $\theta_1 < \theta_2$. If $\lambda_t \le \theta_1$, only the BS with the heaviest traffic load is active. If $\theta_1 < \lambda_t < \theta_2$, the only BS with the lightest load sleeps. Otherwise, all the BSs are active. In this figure, we set $\lambda_{\mathrm{max}} = 7.5, \theta_1 = 3, \theta_2 = 6$. It can be seen that the proposed two-stage DP algorithm performs close to the optimal DP algorithm, and is better than the BS on-off algorithm. Hence, in addition to the BS on-off states, the adaptation of number of active subcarriers and opportunistic sleep ratio further reduces grid power consumption. The threshold-based heuristic sleep algorithm performs better than the non-sleep algorithm when grid power is less than 1550Watt.

\begin{figure}
\centering
\includegraphics[width=4.5in]{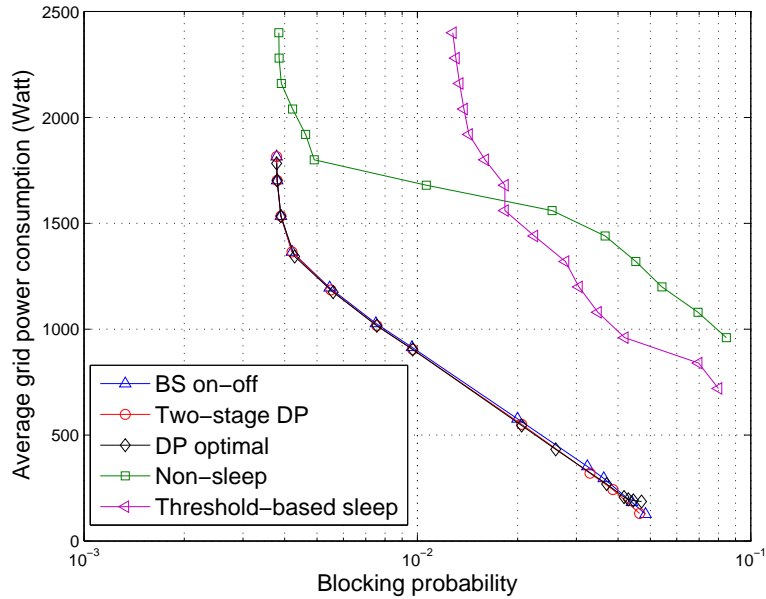}
\caption{Tradeoff curves between outage probability and grid energy consumption for 3-sector case. $K=1, \psi^{(1)}:\psi^{(2)}:\psi^{(3)} = 1:1:1$. The renewable energy profile as in Fig. \ref{fig:traffic} is the same for three BSs.} \label{fig:K1s}
\end{figure}

Fig.~\ref{fig:K1s} depicts the tradeoff curves of different algorithms for the symmetric traffic distribution case. In this figure, it can be seen that the BS on-off algorithm and two-stage DP algorithm are close to each other, which means that the performance improvement by the active subcarrier adaptation and the opportunistic sleep ratio adjustment is not significant. It can be explained as follows. Reducing the number of active subcarriers reduces the available wireless radio resources (enhancing its own blocking probability) on the one hand, but on the other hand reduces the interference to the neighbouring cells (reducing neighbor cells' blocking probability). In the asymmetric traffic distribution scenario, if the number of active subcarriers of low traffic BS is reduced, the effect of interference reduction outweighs that of radio resource reduction as the blocking probability is quite low. As a result, we can adapt the number of active subcarriers to approach the optimal bound. On the contrary, if all the three BSs experience the same traffic conditions, the blocking probability enhancement by the radio resource reduction is larger than the decrease from the interference reduction. Hence, it is better to keep all the subcarriers active.

\begin{figure}
\centering
\includegraphics[width=4.5in]{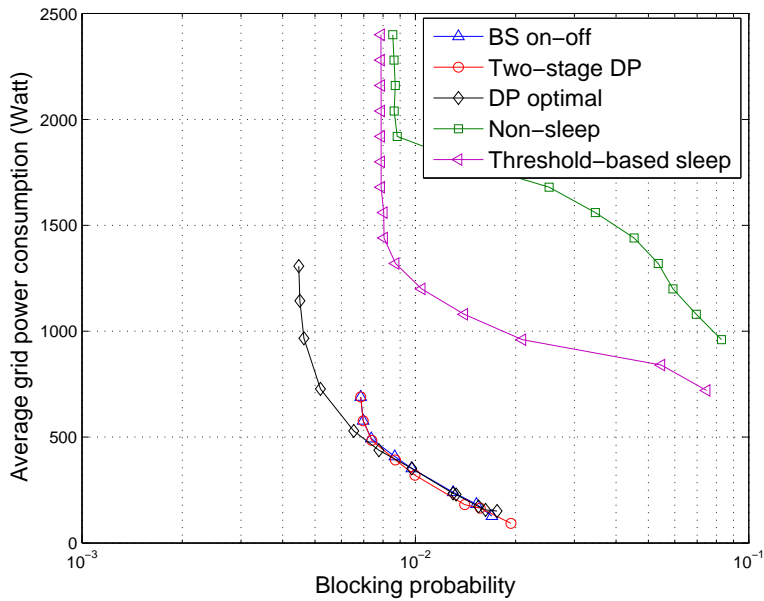}
\caption{Tradeoff curves between outage probability and grid energy consumption for 3-sector case. $K=2, \psi^{(1)}:\psi^{(2)}:\psi^{(3)} = 1:2:3$. The renewable energy profile as in Fig. \ref{fig:traffic} is the same for three BSs.} \label{fig:K2a}
\end{figure}

\begin{figure}
\centering
\includegraphics[width=4.5in]{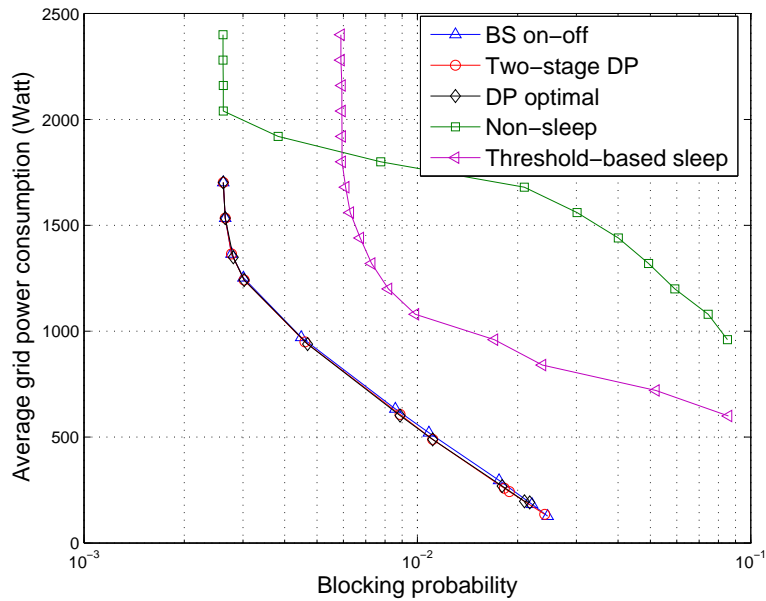}
\caption{Tradeoff curves between outage probability and grid energy consumption for 3-sector case. $K=2, \psi^{(1)}:\psi^{(2)}:\psi^{(3)} = 1:1:1$. The renewable energy profile as in Fig. \ref{fig:traffic} is the same for three BSs.} \label{fig:K2s}
\end{figure}

We also simulate the multiple user class case. Figs.~\ref{fig:K2a} and \ref{fig:K2s} illustrate the tradeoff curves with $K=2$ user classes (rate requirements are $R_1 = 2$Mbps $R_2=0.5$Mbps) for the asymmetric and the symmetric traffic distribution, respectively. We assume $\lambda_{\mathrm{max}} = 12, \theta_1 = 3, \theta_2 = 10$, and each user class occupies half of the traffic. It can be seen that the performance is similar with the $K=1$ case for the symmetric traffic distribution, but is different for the asymmetric traffic distribution. In particular, when a low blocking probability is targeted, the two-stage DP algorithm is not close to the optimal solution any more. From the result, we find that no matter how large $\beta$ is set, we can not activate all the three BSs by the BS on-off algorithm. The reason is that we assume all subcarriers are active in this algorithm, which causes very high interference when all the three BSs are active. However, the optimal DP algorithm can activate all three BSs by jointly optimizing the number of active subcarriers to reduce interference.

\section{Conclusion} \label{sec:concl}
This paper studied the joint optimizing problem of BS sleeping and resource allocation in a long-term point of view using the average network traffic profile and the harvested energy profile. The proposed two-stage DP algorithm is shown to achieve near-optimal performance as long as the first stage BSs' on-off state adaptation achieves the optimal result. In addition, for the symmetric traffic distribution scenario, the results show that the active subcarrier adaptation does not significantly improve performance, which means that we only need to determine BSs' on-off state and the active BSs activate all their subcarriers with sufficient power input. This can greatly reduce the computational complexity for achieving the optimal solution. On the contrary, if the traffic is asymmetrically distributed, active subcarrier adaptation can effectively reduce the interference while guaranteeing radio resources requirement. Hence, the performance can be improved. For the $K=1$ case, if $p_{\mathrm{target}} = 1.25\%$, the two-stage DP algorithm reduced the grid power consumption by about 50\% comparing with the BS on-off algorithm.

\end{document}